\documentstyle[aps,twocolumn]{revtex}
\begin{document}
\draft
\wideabs{
\title
{Enhancement and deterioration of the critical current
by defects in Josephson-junction arrays}
\author{D.-X. Chen, J. J. Moreno, and A. Hernando}
\address
 {Instituto de Magnetismo Aplicado, UCM-RENFE-CSIC,
   28230 Las Rozas, Madrid, Spain}
\author{A. Sanchez}
\address
{Grup d'Electromagnetisme, Departament de F\'\i sica, Universitat
Aut\`onoma de Barcelona\\ 08193 Bellaterra, Barcelona, Catalonia,
Spain}
\maketitle
\begin{abstract}
We study how the addition
of defects in an otherwise uniform Josehpson junction array
modifies
the critical current of the array, by numerically solving the system
of coupled
Josephson equations. Our results confirm the
existence of two different
regimes, depending on the normalized maximum Josephson current of
each junction, $i_{\rm max}$: for large $i_{\rm max}$, the
defects deteriorate partially the critical state of the uniform array
and so decrease the critical current, while for low values of
$i_{\rm max}$, the
defects help the formation of the critical state, which leads to an
enhancement of critical current. The results are discussed in
comparison with the behavior of type-II superconductors, for which
the inclusion of defects always results in an enhancement of the
critical current.
\end{abstract}
}
 %\pacs{PACS numbers: 74.50.+r, 74.60.Ge, 74.60.Jg}

The traditional way of improving the critical current of hard type-II
superconductors has been the introduction of defects, which act as
pinning sites of the vortices. Based on the analogy
between the magnetic properties of superconductors and Josephson
junction (JJ) arrays, it seems reasonable to expect that in the
latter system the presence of defects will yield to an improved total
critical current. This has been the predominant point of view,
starting from the work of
Josephson himself \cite{josep1}. His point of view was followed in
later researches on defects containing JJ's
\cite{pin1,pin4,pin5,pin6}, and in
systematic works on planar JJ arrays, where Josephson vortices (JV's)
and
their depinning were studied (for example, see
\cite{teitel1,lobb1,stroud1,xia,lobb3,teitel3,dang1}).
Moreover, it has become
evident that one of the
limiting factors for the critical current of actual high-$T_c$
superconductors is the presence of a granular structure consisting of
superconducting grains connected by JJ's.
The role played by the defects in JJ arrays with respect to
transport current is therefore a topic of
practical importance as well. Clem introduced the JV pinning
mechanism as the reason for the intergranular critical state (CS)
\cite{clem3}. In this
work, we study this problem by
numerically solving the equations describing the behavior of a
JJ array in which defects are introduced. Our results will
demonstrate that, in contrast with the case of type-II
superconductors, the total critical current in the array decreases
when adding
defects, except in the regime of low Josephson current, for which the
inclusion of defects may help the formation of a CS which
does not originally appear in the uniform JJ array.

In order to avoid complicated demagnetizing effects, the studied JJ
array
consists of a number of infinitely long superconducting grains along
the $z$ axis.
Its $xy$-plane cross-section forms an $N_x\times N_y$ square lattice
of parameter $a$, each grain being centered at $(i,j)$, with
$i=1,2,\ldots, N_x$ and $j=1,2,\ldots, N_y$.
Every pair of nearest grains is weakly linked by a (when looking at
it
along the $z$ axis) short JJ,
which has a line density of the maximum DC Josephson current $I_{\rm
max}$
(A/m) and a line specific resistance $R$ ($\Omega$m), the line being
along the $z$ dimension.  The capacitance effects are neglected for
studying a slow process.
Four nearest grains with an enclosed
void form a square cell, the effective void area being $A_{\rm v}$
(m$^2$).
We name the gauge-invariant phase differences (GIPD's) for JJ's along
the
$x$ and
$y$ axes $\theta^x_{ij}$ and $\theta^y_{ij}$, respectively.

When a
transport current with line density $I_{\rm array}$ is uniformly
applied in the $x$
direction, or a field $H$ is uniformly applied in the $z$ direction,
a set
of differential
equations for the $2N_xN_y-N_x-N_y$
GIPD's of all JJ's of the JJ array is built up
based on the DC and AC Josephson equations and the Amp\`ere and the
Ohm
laws as

\begin{eqnarray}%3
\label{main}
{{\rm d}\theta^x_{ij}\over {\rm d}t^*}&=&-2\pi h_0
-\theta^x_{ij}+\theta^x_{i,j+1}+\theta^y_{ij}-\theta^y_{i+1,j}
\nonumber \\
&-&2\pi i_{\rm max}\sin\theta^x_{ij}\;\;(1\leq i\leq N_x-1,j=1),
\nonumber \\
{{\rm d}\theta^x_{ij}\over
{\rm d}t^*}&=&\theta^x_{i,j-1}-2\theta^x_{ij}+\theta^x_{i,j+1}
+\theta^y_{ij}-\theta^y_{i,j-1}+\theta^y_{i+1,j-1}
\nonumber \\&-&\theta^y_{i+1,j}-2\pi i_{\rm
max}\sin\theta^x_{ij}\nonumber\\
&&\hspace{2cm}(1\leq i\leq N_x-1, 2\leq j\leq N_y-1),
\nonumber \\
{{\rm d}\theta^x_{ij}\over {\rm d}t^*}&=&2\pi h_{N_y}
+\theta^x_{i,j-1}-\theta^x_{ij}-\theta^y_{i,j-1}+\theta^y_{i+1,j-1}
\nonumber \\&-&2\pi i_{\rm max}\sin\theta^x_{ij}\;\;(1\leq i\leq
N_x-1,j=N_y),
\nonumber \\
{{\rm d}\theta^y_{ij}\over {\rm d}t^*}&=&2\pi h_j
-\theta^y_{ij}+\theta^y_{i+1,j}+\theta^x_{ij}-\theta^x_{i,j+1}
\nonumber \\&-&2\pi i_{\rm max}\sin\theta^y_{ij}\;\;(i=1,1\leq j\leq
N_y-1),
\nonumber \\
{{\rm d}\theta^y_{ij}\over
{\rm d}t^*}&=&\theta^y_{i-1,j}-2\theta^y_{ij}+\theta^y_{i+1,j}
+\theta^x_{ij}-\theta^x_{i-1,j}+\theta^x_{i-1,j+1}
\nonumber \\&-&\theta^x_{i,j+1}
-2\pi i_{\rm max}\sin\theta^y_{ij}\nonumber \\
&&\hspace{2cm}(2\leq i\leq N_x-1,1\leq j\leq N_y-1),
\nonumber \\
{{\rm d}\theta^y_{ij}\over {\rm d}t^*}&=&-2\pi h_j
+\theta^y_{i-1,j}-\theta^y_{ij}-\theta^x_{i-1,j}+\theta^x_{i-1,j+1}
\nonumber \\&-&2\pi i_{\rm max}\sin\theta^y_{ij}\nonumber \\
&&\hspace{2cm}(i=N_x,1\leq j\leq N_y-1).
\end{eqnarray}
In these equations, $t^*$ is the normalized time $t$ to the nominal
time constant $\tau$ of one cell, $t^*=t/\tau=tR/\mu_0A_{\rm v}$,
and $h_j$ and $i_{\rm max}$ are the boundary field $H_j$ and
$I_{\rm max}$ normalized to $\Phi_0/\mu_0A_{\rm v}$.
Correspondingly, the normalization of $I_{\rm array}$
into $i_{\rm array}$ can be made in the same way as $I_{\rm max}$.
We have $h_j=(j/N_y-1/2)i_{\rm array}$ in the transport case and
$h_j=h$ in the magnetic case.
The field $H_{ij}$ in the $(i, j)$
void ($i$ and $j$ are defined as those for the left-and-down adjacent
grain) produced by DC Josephson currents
 normalized to $\Phi_0/\mu_0A_{\rm v}$ is calculated by
\begin{equation}%6
\label{coupling}
h_{ij}=(\theta^y_{ij}-\theta^y_{i+1,j}+\theta^x_{i,j+1}-
\theta^x_{ij})/2\pi,
\end{equation}
where we have assumed the phase of the order parameter to be zero in
each grain.

It is known \cite{greene,doro} that a uniform slablike JJ array
\cite{footnote}  can be in
the CS when $i_{\max}$ is
large enough, while for lower values of $i_{\max}$ it presents a
JV state.
Such an evolution in 1D JJ array from the JV state to the CS has been
studied mathematically
by Greene in terms of a standard mapping technique.  His conclusion
is that
a stochastic transition occurs at $i_{\rm
max}=i_{\rm max}^*=0.9716\cdots/2\pi\approx 0.155$
\cite{greene,doro}.

The transition from a JV state to a CS occurs in our
(mathematically 2D) JJ array as well. To show this transition, we
analyze the field profiles in a JJ array of $N_x=N_y=25$ in the
saturated remanence state. The profiles are calculated using a
Runge-Kutta method as follows.
We start from $\{\theta^{x,y}_{ij}=0\}$, step $h$ from 0 to 100 and
calculate the fully relaxed $\{h_{ij}\}$, and then step $h$ down to 0
and calculate again the fully relaxed $\{h_{ij}\}$, which is our
solution.

As seen in Fig. 1, the profiles of uniform JJ arrays with $i_{\rm
max}=0.01,0.05,$ and
0.1 show JV states with 8, 24, and 64 JV's of roughly equal heights.
For
$i_{\rm max}=0.17$, there is a central peak remarkably higher than
the
others, indicating a transition from the JV state to the CS.
When $i_{\rm max}\geq 0.4$, the entire profile appears to
be a single
peak of pyramid shape (as shown in the figure for $i_{\rm max}=1$),
which is the distinctive characteristic of the Bean
CS, as in hard superconductors, where
the penetrated supercurrents flow with a density $J$
equal to the critical-current density $J_{\rm c}$ \cite{bean1}.

Let us now study the effect of introducing defects in the array
by changing $i_{\rm max}$ of JJ's connecting certain
defect grains into $i_{\rm max,d}=ci_{\rm max}$, where $0<c\neq 1$.
First we consider the case for
low values of $i_{\max}$, for which the
uniform JJ array does not show a CS. In order to be able to check
the validity of the solutions by their symmetry, we choose
symmetrically and uniformly distributed defects shown in Fig. 2(a) with
$c=0.1$.
The trend observed in the calculated results shown in Fig. 1 is
clear:
defects help the formation of the CS.
For the nonuniform JJ array of $i_{\rm
max}=0.01$, there is a 12-peak
structure with eight outer peaks lower than the inner four.  When
$i_{\rm
max}$ is increased to 0.05, the profile appears to be a single round
peak, indicating a transition from the JV state to the CS.  The
profiles
for $i_{\rm max}=0.1$ and 0.17 are already of an overall pyramid
shape, typical
for the Bean CS, in clear contrast to the results for the uniform
JJ array. We confirm therefore that the inclusion of defects
in the array with low values of $i_{\rm
max}$ results in the enhancement and even induction of the CS.

After this conclusion, an important question arises as: Is
this behavior general for all values of $i_{\rm max}$? The previous
result of the decreasing of the trapped flux for $i_{\rm max}=1$ [as
seen in Fig. 1(i)]
already gives us a hint of the actual striking behavior:
for such values of
$i_{\max}$ that the uniform array develops a good CS, the
addition of defects results in a decrease of critical current.

To further confirm the above statement we set two square defect
clusters shown in Fig.~2(b).
Our aim is to calculate the line density of the critical current of
the JJ array,
$i_{\rm c,\rm array}$,
namely, the maximum $i_{\rm array}$
at which a static solution can be obtained.  The
reference is chosen as the uniform ($c=1$) JJ array of $i_{\rm
max}=1$.
We start from a zero-field cooled state at $t^*=0$ with
 $\{\theta_{ij}^{x,y}=0\}$, and change $i_{\rm array}$
between 0 and 25 (the maximum possible value for $i_{\rm max}=1$)
to calculate the fully relaxed GIPD's that corresponds to $i_{\rm
c,array}$.

The calculated $i_{\rm c,\rm array}$ as a function of $c$ is given
in Fig.~3.  We can see that the presence of defects makes
$i_{\rm c,\rm array}$ smaller in general.  It equals 24 when
$0.995\leq
c\leq 1.519$, and decreases stepwise if $c$ is out of this range.
The reasons for this are twofold.  (i) Since $i_{\rm array}$
flows in the horizontal $x$ direction
through JJ's of $i=1$ to 25 in series, it is limited  by the ideal critical
current $i_{\rm c,\rm array, ideal}$ defined as the smallest
sum of $i_{\rm max}$ and $i_{\rm max,d}$ of all the horizontal JJ's for
each $i$th vertical cross-section.  For the array in Fig. 2(b),
$i_{\rm c,\rm array, ideal}=25$ if $c\geq 1$ and
$i_{\rm c,\rm array, ideal}=19+6c$ if $c<1$.  (ii) The phase coupling
related to Eq. (2) makes $i_{\rm c,array}<i_{\rm c,\rm array,
ideal}$, so that the calculated $i_{\rm c,array}<25$ if $c\geq 1$ and
$i_{\rm c,array}<19+6c$ if $c<1$.  The effect of the
defects on the critical current of the nonuniform array can be
observed in greater detail from the calculated current distributions in
the array for different values of $c$ presented in \cite{reviewjja}.

To sum up, we have shown that the addition of defects in a
JJ array helps to settle a CS in the
system if the normalized JJ current $i_{\rm
max}$ is small. However if $i_{\rm max}$ is large enough, that is for
the cases in which the uniform array is in
a good CS, the introduction of defects deteriorate the
critical current of the array.
The latter feature is in contrast with the popular view that the
CS arises
from JV pinning. From these results, it is
imperative to conclude that the origin of the CS in
JJ arrays is a phenomenon more complex than a
JV pinning analogous to the Abrikosov vortex pinning in hard
type-II superconductors.

We thank Ministerio de Ciencia
y Tecnolog\'\i a project number BFM2000-0001, and CIRIT project
number 1999SGR00340 for financial support.

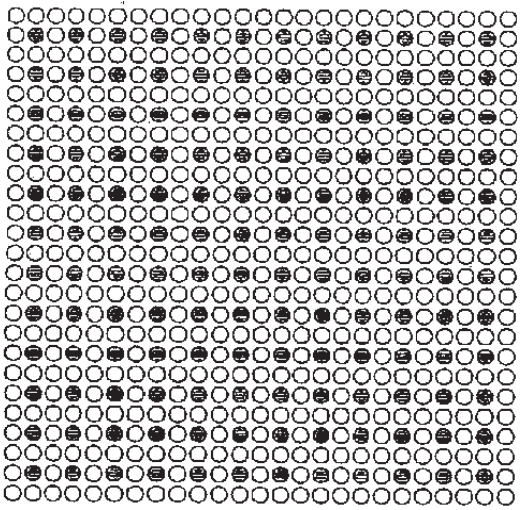
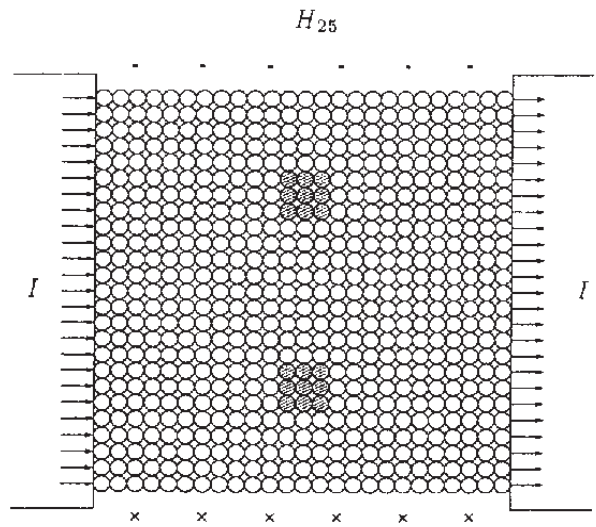
\begin{figure}
\caption
{Trapped field profiles of $25\times 25$ square-columnar JJ arrays at
saturated remanence. (a), (c), (e), (g), and (i) are for uniform
JJ arrays; (b),
(d), (f), (h), and (j) are for nonuniform JJ arrays with $i_{\rm
max,d}=i_{\rm
max}/10$.  Relative $h$ scale factors $1/i_{\rm max}$ are used.}
\end{figure}

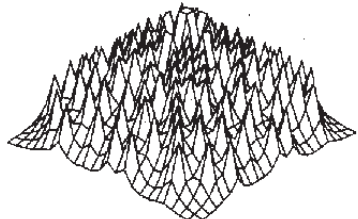
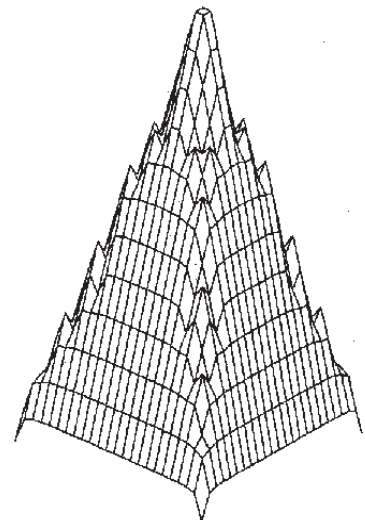
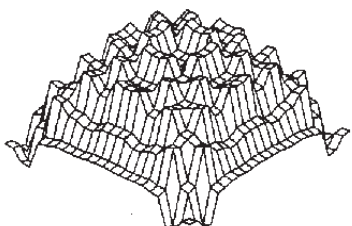
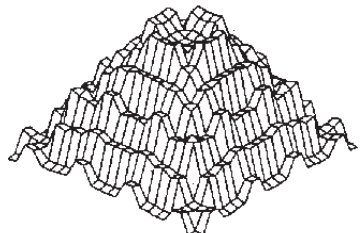
\begin{figure}
\caption
{The $xy$ cross-section of the studied JJ arrays. The defect grains
are
marked by shadow. (a) and (b) are
chosen for the cases of small and
large values of $i_{\rm max}$, respectively. }
\end{figure}

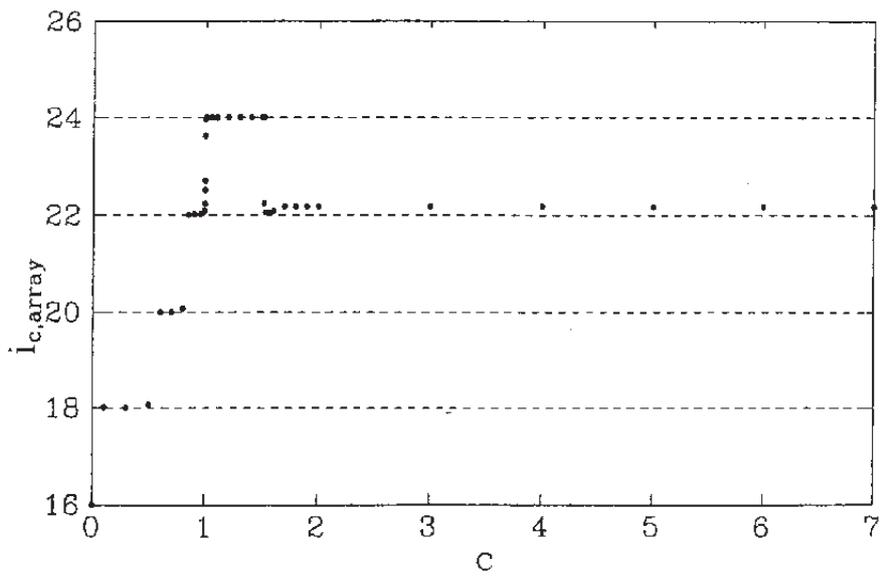
\begin{figure}
\caption
{The JJ critical current $i_{\rm c,\rm array}$ as a function of $c$.}
\end{figure}


\begin{references}
\bibitem{footnote} The slablike is defined as a JJ array studied in
the
present work with $N_y=\infty$, so each row of grains or JJ's
should be in the same state when the JJ array is placed in a uniform
field
in the $z$ direction; it is the analogous to a superconducting slab.
\end{references}
\end{document}